\documentclass[11pt]{article}
\usepackage{latexsym}

\usepackage{amssymb}
\usepackage{amsmath}

 \usepackage{epsfig}

\numberwithin{equation}{section}

\newcommand{\be}{\begin{equation}}
\newcommand{\ee}{\end{equation}}
\newcommand{\ba}{\begin{eqnarray}}
\newcommand{\ea}{\end{eqnarray}}
\newcommand{\baa}{\begin{array}}
\newcommand{\eaa}{\end{array}}
\newcommand{\bi}{\begin{itemize}}
\newcommand{\ei}{\end{itemize}}
\newcommand{\nn}{\nonumber \\}
\newcommand{\nr}[1]{(\ref{#1})}

\newcommand{\lt}{{\tilde\lambda}}
\newcommand{\ltz}{{\tilde\lambda}^{(0)}}
\newcommand{\la}[1]{\label{#1}}
\newcommand{\bmu}{\bar\mu}
\newcommand{\rmi}[1]{{\mbox{\scriptsize #1}}}
\newcommand{\fr}[2]{{\frac{#1}{#2}\,}}
\newcommand{\fra}[2]{\textstyle{\frac{#1}{#2}\,}}  
\newcommand{\mn}{{\mu\nu}}
\newcommand{\bfx}{{\bf x}}

\newcommand{\edoc}{\end{document}}

\def\CL{{\cal L}}

\def\gsim{\raise0.3ex\hbox{$>$\kern-0.75em\raise-1.1ex\hbox{$\sim$}}}
\def\lsim{\raise0.3ex\hbox{$<$\kern-0.75em\raise-1.1ex\hbox{$\sim$}}}

\begin{document}

\begin{titlepage}
\begin{flushright}
HIP-2009-13/TH\\
Revised 10 September 2009\\
\end{flushright}
\begin{centering}
\vfill

{\Large{\bf Spatial string tension of finite temperature
QCD matter in gauge/gravity duality}}

\vspace{0.8cm}

\renewcommand{\thefootnote}{\fnsymbol{footnote}}

J. Alanen$^{\rm a,b}$\footnote{janne.alanen@helsinki.fi},
K. Kajantie$^{\rm a,b}$\footnote{keijo.kajantie@helsinki.fi},
V. Suur-Uski$^{\rm a,b}$\footnote{ville.suur-uski@helsinki.fi}

\setcounter{footnote}{0}

\vspace{0.8cm}

{\em $^{\rm a}$%
Department of Physics, P.O.Box 64, FI-00014 University of Helsinki,
Finland\\}
{\em $^{\rm b}$%
Helsinki Institute of Physics, P.O.Box 64, FI-00014 University of
Helsinki, Finland\\}

\vspace*{0.8cm}

\end{centering}

\noindent
The computation of the spatial string tension of finite temperature QCD is
discussed in QCD and in a gravity+scalar model of gauge/gravity duality.

\vfill \noindent

%

\vspace*{1cm}

\noindent

September 2009

\vfill

\end{titlepage}

\section{Introduction}
The spatial string tension $\sigma_s(T)$ in finite temperature QCD is determined
with lattice Monte Carlo techniques \cite{manousakis,bali,boyd,liddle,cheng} by studying
in asymmetric $N_t\ll N_s$ lattices
expectation values of large Wilson loops with both sides in spatial directions. It is
a nonperturbative quantity and thus not analytically calculable. However, if it is determined
numerically in 3 dimensional (3d) SU(3) gauge theory \cite{teper}, one can use the well established
equations of perturbative high temperature dimensional reduction to compute it in
full 4d finite $T$ QCD and compare it with the lattice result \cite{laine,lainejork1}.
The agreement is surprisingly good.

Since $\sigma_s(T)$ is so well under control in QCD, it is a good laboratory for
testing various models for AdS/QCD \cite{andreev1,andreev2,andreev3}. The heart
of QCD computations is conformal invariance breaking
and running of the coupling constant while the heart of well established AdS/CFT duality
is conformal invariance of CFT. The essential part of AdS/QCD models is thus how
conformal invariance breaking is modelled. We shall in this article apply the AdS/QCD
model of Kiritsis and coworkers \cite{kiri1,kiri2,kiri3,kiri4},
approximating it in a form which permits an analytic solution.  

\section{QCD discussion}\label{sect2}
A first-principle method for determining the spatial string tension in hot QCD matter
is to measure rectangular Wilson loops of size $X\times Y$ in the $(x,y)$ plane.
A potential $V(L)$ is then defined by:
\be
V(L)=-\lim_{Y\to\infty}{1\over Y}\log W(L,Y)
\la{pot}
\ee
and then finding $\sigma_s$ from the large-$L$ behavior
$$
\sigma_s(T)=\lim_{L\to\infty} {V(L)\over L}.
$$
Measurements are done with full 4d QCD action on a finite $T$ lattice
$aN_t=1/T\ll aN_s$ and they
can be done for pure SU(3) Yang-Mills theory or for QCD with dynamical quarks.

There is also a natural way to interpret the result and to derive it by splitting the problem
in an analytic and a numerical nonperturbative part. The spatial string tension lives
in the 3d spatial space and one can also determine it for 3d SU(3) Yang-Mills theory. Since
the only dimensionful parameter in this theory is its coupling constant
$g_M^2$ of dimension mass, one knows that the string tension
must be a number times $g_M^4$. In fact \cite{teper,teper1},
\be
\sqrt{\sigma_s}=0.553(1)g_M^2. \label{sig3d}
\ee
On a schematic level, one knows that $g_M^2=g^2(T)T$, where the 4d coupling constant
$1/g^2(T)$ has a simple
logarithmic expansion. Thus it is natural to choose to plot
\be
{T\over\sqrt{\sigma_s}}= {1\over 0.553}\,{1\over g^2(T)}=
1.81\cdot{22N_c\over 3(4\pi)^2}\left[\log{T\over\Lambda_\sigma}+{51\over121}
\log\left(2\log{T\over\Lambda_\sigma}\right)\right],
\la{tovers}
\ee
where the standard 2-loop expansion
is inserted (with $N_f=0$). The scale $\Lambda_\sigma$ is unknown but can be fitted
to the data. The data for finite $T$ SU(3) Yang-Mills theory \cite{boyd}
is as shown below in Figs.~\ref{fig3} and \ref{fig4}.

To obtain a controlled theoretical result one at least has to compute the $\Lambda_\sigma$
in \nr{tovers} and not fit it to data. To do this
one can start from the nonperturbative
magnetic sector 3d number \nr{sig3d} and work out backwards, first to the
electric sector of hot QCD by restoring the electric mass $m_\rmi{E}\sim gT$ by
the 2-loop relation \cite{giovannangeli}
\be
 g_\rmi{M}^2 = g_\rmi{E}^2
 \; \biggl[
  1 - \frac{1}{48} \frac{g_\rmi{E}^2 N_c}{\pi m_\rmi{E}}
    - \frac{17}{4608}
  \biggl( \frac{g_\rmi{E}^2 N_c}{\pi m_\rmi{E}} \biggr)^2+
  {\cal O}\biggl({g_E^2N_c\over \pi m_E}\biggr)^3
 \biggr]
 \;, \la{gMgE}
\ee
and finally the hard scale $\pi T$ by the 2-loop relations \cite{laine} ($\beta_0=-22N_c/3,\,
\beta_1=-68N_c^2/3$)
\ba
 g_\rmi{E}^2/T & = &
 g^2(\bmu) +
 \frac{g^4(\bmu)}{(4\pi)^2}
 \Bigl[
 -\beta_0 \ln  \biggl( \frac{\bmu e^{\gamma_\rmi{E}}}{4\pi T} \biggr)
 + \fr13 N_c
 \Bigr] \nn &&+\,\,
 \frac{g^6(\bmu)}{(4\pi)^4}
  \biggl\{
  -\beta_1 \ln  \biggl( \frac{\bmu e^{\gamma_\rmi{E}}}{4\pi T} \biggr)
 + \biggl[
 \beta_0 \ln  \biggl( \frac{\bmu e^{\gamma_\rmi{E}}}{4\pi T} \biggr)
 - \fr13 N_c
 \biggr]^2
 \nn &   &-\,\,
 \frac{1}{18}
 N_c^2 \Bigl[ -341 + 20 \zeta(3) \Bigr]
  \biggr\} + \mathcal{O}(g^8)\nn
  &=& g^2(\bmu_\rmi{op})+\frac{g^6(\bmu_\rmi{op})}{(4\pi)^4}{1\over198}N_c^2[3547-220\zeta(3)]
 \;, 
 \la{gE1}
\ea
\ba
 m_\rmi{E}^2/T^2 &=&
  g^2(\bmu)
   {1\over3}N_c
  +
 \frac{g^4(\bmu)}{(4\pi)^2}
   N_c^2 \left(\frac{22}{9} \ln\frac{\bmu e^{\gamma_\rmi{E}}}{4 \pi T} + \fr59  \right)
 + \mathcal{O}(g^6)\nn
 &=&g^2(\bmu_\rmi{op})
   {1\over3}N_c
  +
 \frac{g^4(\bmu_\rmi{op})}{(4\pi)^2}
   {4\over9}N_c^2
 \;, \la{massE}
\ea

\ba
{1\over g^2(\bmu_\rmi{op})}&=&
{22N_c\over 3(4\pi)^2}\left[\log{7.753T\over T_c}+{51\over121}
\log\left(2\log{7.753T\over T_c}\right)\right].
\la{g2op}
\ea
Here $g^2(\bmu)$ is the $\overline{\rm MS}$ running coupling in which
we, for concreteness, optimized the scale $\bmu$ so that the $g^4$ term in \nr{gE1} vanishes,
i.e., $\bmu_\rmi{op}=4\pi e^{-\gamma_\rmi{E}-1/22}T=6.742\,T$.
A thorough analysis of the optimisation scale dependence is given in \cite{laine}.
Noting that $T_c=1.15\Lambda_{\overline{\rm MS}}$ \cite{laine}
the logarithmic factors in $1/g^2(\bmu_\rmi{op})$ could be written as $\log(7.753\,T/T_c)$, i.e.,
$\Lambda_\sigma$ in \nr{tovers} has been evaluated to be $\Lambda_\sigma=T_c/7.753$.

The reliable QCD prediction then is obtained by inserting to \nr{sig3d} $g_M^2$ from \nr{gMgE}
with $g_E^2$ and $m_E^2$ as given by \nr{gE1} and \nr{massE}\footnote{Note that in a
3d gauge theory the couplings are scale independent while
$m^2$ has both a linear 1-loop and a logarithmic 2-loop divergence. It so happens that
the $g^4$-coefficient of the logarithmic divergence cancels for $m_E^2$ \cite{perturbative} and
$m_E^2$ above is scale independent to the order shown,
$\bmu\,\partial m_E^2/\partial\bmu={\cal O}(g^6)$.}.
The prediction is shown as the continuous curve
plotted over the range $T_c<T<20\,T_c$ in Figs.~\ref{fig3} and \ref{fig4} below.
The lattice data, shown in
the same figures, extends over the range $T_c<T<4.5\,T_c$.

One also observes that the correction terms in \nr{gMgE} and the $g^6$ term in \nr{gE1} almost
cancel each other so that to a good accuracy the result is just given by the
2-loop expression in \nr{tovers}.

The QCD computation is an entirely controlled
perturbative computation, the nonperturbative part is isolated in the number \nr{sig3d}. It
works surprisingly well, perhaps even too well. The computation is, namely, based on
integrating out the large scales $\pi T$ and $gT$ and one does not expect it to work
down to $T_c$. For the pressure a similar computation shows clear deviations from the
lattice results for $T$ below $\sim 4T_c$ \cite{hietanen}.

Clearly, to assess the reliability of the result it would be useful to further compute the 3-loop
contribution -- a formidable task.

\section{AdS/QCD model}
Consider then the computation of the spatial string tension in AdS/QCD models. In bottom-up
models
conformal invariance can be broken by putting by hand some structure in the extra dimension
$z$: a hard wall or some soft function of the type $\exp(c z^2)$ \cite{kty}.
We shall use a model \cite{kiri3} the idea of which is to add a scalar field and
to create this scale dynamically via the equations of motion. Related work on vacuum
potentials is in \cite{zeng}.

\subsection{Defining the AdS/QCD model}          
\begin{figure}[!tb]
\begin{center}

\vspace{-0.8cm}
\includegraphics[width=0.5\textwidth]{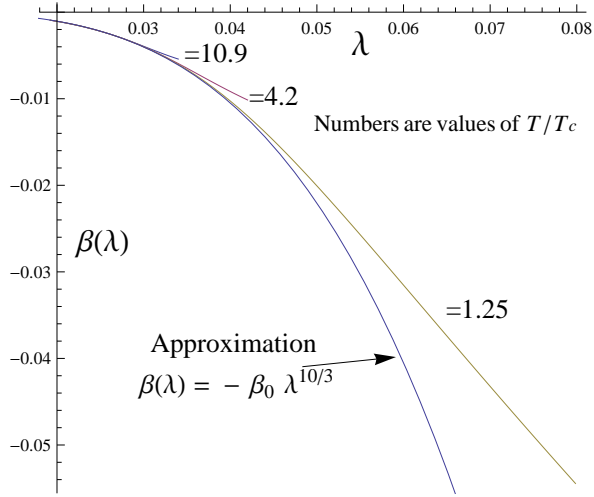}
\end{center}

\caption{\small The "thermal" $\beta$ function, defined by the relation
$\beta(\lambda)=b d\lambda/db$, from numerics using the full form \nr{pot} of the potential.
Since there is a horizon at some $z=z_h$, the functions
$b(z),\lambda(z)$ terminate at $z_h$ and thus also
$\beta(\lambda)$. This $z_h$ corresponds to some $T/T_c$, marked on the figure.
The true "vacuum" beta function is approached when $z_h\to\infty$.
The potential \nr{pot} is so constructed that for large $\lambda$,
$\beta=-\fra32\lambda(1+3/(8\log\lambda))$, as required by confinement. For the $\beta$ function
at $T=1.25T_c$ the asymptotic slope is $-1.1\lambda$.
At $T=0.96T_c$ the slope is $-1.47\lambda$. The QCD universal beta function
$-b_0\lambda^2-b_1\lambda^3$ is only approached at $\lambda=0.006$.
}
\label{fig_thermalbeta}
\end{figure}

One starts from the gravity + scalar action (in the Einstein frame and in standard notation)
\be
S={1\over16\pi G_5}\left\{\int d^5x\,\sqrt{-g}\left[R-\fra43(\partial_\mu\phi)^2+V(\phi)\right]
-2\int d^4x\,\sqrt{-\gamma}K\right\}.
\la{Eframeaction}
\ee
or, in the string frame, writing
\be
g^s_{\mn}=e^{\fr43\phi}g^E_{\mn},
\la{stringeinstein}
\ee
\be
S={1\over16\pi G_5}\left\{\int d^5x\,\sqrt{-g}e^{-2\phi}\left[R+4(\partial_\mu\phi)^2+
e^{-\fr43\phi}V(\phi)\right]-
2\int d^4x\,\sqrt{-\gamma}K\right\}.
\ee
One now assumes a metric ansatz
\be
ds^2=b^2(z)\left[-f(z)dt^2+d\bfx^2+{dz^2\over f(z)}\right].
\la{ansatz}
\ee
The four functions $b(z), f(z)$ in the metric, the scalar field $\phi(z)$ and the
potential $V(\phi(z))$ are then
determined as the solutions of the three field equations following from \nr{Eframeaction}:
\ba
&&6{\dot b^2\over b^2}+3{\ddot b\over b}+3{\dot b\over b}{\dot f\over f}={b^2\over f}V(\phi),\label{eq1}\\
&& 6{\dot b^2\over b^2}-3{\ddot b\over b}={\fra43} \dot\phi^2,\label{eq2}\\
&&{\ddot f\over \dot f}+3{\dot b\over b}=0,\label{eq3}
\ea
($\dot b\equiv b'(z)$, etc.) and from a fourth equation,
\be
\beta(\lambda)=b{d\lambda\over db},\quad \lambda(z)  = e^{\phi(z)}\sim N_c g^2,
\label{crucial}
\ee
where $\beta(\lambda)$ is the beta function of the field theory one is seeking the
gravity dual for.
This is the crucial assumption of the model. The logic here is that
the energy scale of the coupling is identified by $1/z$, so that $z\to0$ ($z\to\infty$)
corresponds to the UV (IR). Any similar monotonic function would do and actually in
the definition \nr{crucial} one uses $E\sim b(z)/\CL$. Eq.\nr{crucial} in the context of
a black hole horizon at some $z=z_h$ is discussed below (see Fig. \ref{fig_thermalbeta}).

For given $\beta(\lambda)$ one can from the fourth equation solve $b=b(\lambda)$, then from the
third equation $f=f(\lambda)$ and finally from the second equation $\lambda=\lambda(z)$ so that
also $b(z)$ and $f(z)$ are determined, in terms of a number of integration constants.
The role of the first equation then simply is to fix the
scalar potential. Defining
\be
W= -\dot b/b^2,\qquad \dot b\equiv db/dz
\ee
the answer simply is
\be
V(\phi)=12fW^2\left[1-\left({\beta\over3\lambda}\right)^2\right]-3{\dot f\over b}W,
\label{V}
\ee
where from the $f$-independent Eq. \nr{eq2}
(which can be written in the form $b\dot W=\fra49\dot\phi^2$) and \nr{crucial}
\be
W(\lambda)=W(0)\exp\left(-\fra49\int_0^\lambda d\bar\lambda{\beta(\bar\lambda)\over\bar\lambda^2}\right),
\qquad W(0)={1\over\CL}.
\label{W}
\ee
The normalisation $W(0)$ follows from the requirement that the boundary be asymptotically AdS,
$V(0)=12/\CL^2$.

In spite of appearances, the potential in Eq.\nr{V} is expressible in terms of $\lambda$
only. A thorough discussion of the choice
of the potential has been given in \cite{kiri1}-\cite{kiri4}. The outcome is that SU(3) thermodynamics
is well described by the above model with the potential\footnote{The numerical
computation includes in \nr{pot} also two more terms
which enforce QCD asymptotic freedom in the UV. These terms are totally irrelevant for
how the model in \cite{kiri4} describes thermodynamics of hot SU(3) matter. In fact, they dominate
for $\lambda<0.006$ which, using \nr{laparam}, corresponds to $T\gsim 10^{58}T_c$}
\be
V(\phi)={12\over\CL^2}\left\{1+V_1\lambda^{4/3}[\log(1+V_3\lambda^2)]^{1/2}\right\}.
\label{pot}
\ee
To find predictions of the model one must solve the Einstein equations numerically, as
discussed in detail on \cite{kiri3,kiri4}. For concreteness,
the parameter values and the normalisation of $\lambda(z)$ used there are
\be
V_1=14.3,\quad V_3=170.4,\quad \lambda(z=1)=0.0242254. \label{params}
\ee

To supplement numerical computations it is very useful to have an analytic approximation.
Clearly in the UV $\lambda\sim g^2$ is small, but the outcome of the numerical computation
is that even at $T=T_c$ $\lambda$ is small, in fact $\approx0.16$. Assuming $\lambda$ is
small, $\lambda <1/\sqrt{V_3}\approx0.08$, one has
\be
V(\phi)\approx
{12\over\CL^2}\left(1+V_1\sqrt{V_3}\lambda^{7/3}\right).
\label{potappro}
\ee
One observes from \nr{V} that a $\beta$ function
\be
\beta(\lambda)=-\beta_0\lambda^q
\label{betaappro}
\ee
leads to
\be
V(\phi)={12\over\CL^2}\left(1+{8\beta_0\over 9(q-1)}\lambda^{q-1}+{\cal O}(\lambda^{2(q-1)})\right).
\ee
Thus the approximate form \nr{betaappro} with
\be
\beta_0=\fra{21}8V_1\sqrt{V_3}=488.8,\qquad q=\fra{10}3,
\la{paramvalues}
\ee
gives the approximate potential in \nr{potappro}. We shall base our analysis on this approximate
beta function.

Analytical solutions for the metric and the scalar for the model beta function \nr{betaappro}
are given in the Appendix. Both small $z,\,\lambda$ or large $T/T_c$ expansions and
bulk thermodynamics can be derived from
them. Physically, they are accurate enough for $T\gsim 1.5T_c$, but clearly the full
potential \nr{pot} is required to describe the phase transition region.

\begin{figure}[!tb]
\begin{center}

\vspace{0.8cm}
\includegraphics[width=0.5\textwidth]{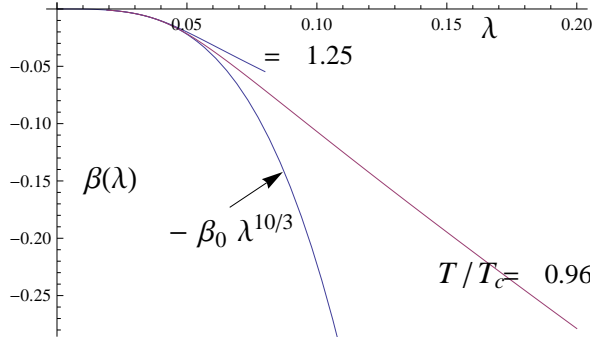}
\end{center}

\caption{\small Same as previous figure but at smaller $T$.
For $T=0.96T_c$ the slope is $-1.47\lambda$, already very close to the
asymptotic large $\lambda$ value
$-3/2\cdot\lambda$. Even here explicitly $\beta(\lambda_h)>-\fra32\lambda_h$.
}
\label{fig_thermalbetasmallT}
\end{figure}

A numerical solution
\cite{kiri4,nitti} of Eqs.\nr{eq1}\,-\,\nr{crucial} gives for $z\ge0$ the functions
$b(z),\,f(z),\,\lambda(z)$ and the value of $T=-\dot f(z_h)/(4\pi)$ for each value of
$\lambda(z)$ at the horizon: $\lambda_h=\lambda(z_h)$. Some properties of the functions
computed are as follows ("numerics" refers to \cite{kiri4}):
\begin{enumerate}
\item $zb(z)\to\CL$ when $z\to0$ and, related to this, $f(0)=1$. In numerics $\CL=1$.

\item There is a horizon at $z=z_h$, $f(z_h)=0$. In numerics there is a different
value of $z_h$ for each value of $T$.

\item $\lambda(z)$ increases monotonically with $z$ from $\lambda(0)=0$ so that
the leading term at small $z$ is
\be
\lambda(z)={1\over \left[-(q-1)\beta_0\log(\Lambda z)\right]^{1/(q-1)}}=
{1\over \left[1140.62\log(174.67/z)\right]^{3/7}},
\la{laparam}
\ee
where $\Lambda$ (=1/174 in numerics) is a constant fixed by the normalisation \nr{params}.

\item The entropy density of the boundary theory is given by
\be
s=s(T)={S\over V_3}={1\over4G_5}b^3(z_h).
\ee
The determination of full bulk thermodynamics and, in particular, of the phase transition
temperature $T_c$ ($\pi T_c=1/198=0.88\Lambda$ in numerics)
between the two phases (the high (low) $T$ phase with $f\not=1$ ($f=1$) in
\nr{ansatz}) is discussed in \cite{kiri4}. 

\item The numerically computed
solutions reproduce the vacuum beta function via the relation $bd\lambda/db=\beta(\lambda)$ only
when $z_h\to\infty$. In fact, if there is a
horizon at some $z=z_h$, the functions $b(z),\lambda(z)$
terminate at $z_h$ and thus also $\beta(\lambda)$.
The outcome of a computation of a "thermal beta function"
defined by the relation $\beta(\lambda)=b d\lambda/db$ is shown in Figs. \ref{fig_thermalbeta}
and \ref{fig_thermalbetasmallT}.
One should emphasize that these are computed using the full potential \nr{pot}.
One also sees how well the approximate beta function \nr{betaappro}
reproduces the beta function corresponding to the
potential \nr{pot}.

\end{enumerate}

\vspace{0.8cm}

\subsection{Deriving the spatial string tension}
The equations giving the extremal string configurations and the corresponding potential are
well known. There are two cases: the spatial string tension at finite $T$ is determined
from a space-space $x,y$ Wilson loop, the usual string tension from a time-space
$t,x$ Wilson loop at $T=0$.
We shall summarize the equations separately for these two cases. The end points of the
string are always at $x=\pm L/2$ and the string hangs in the fifth dimension so that
the maximum depth it reaches at $x=0$ is $z_*<z_h$. We abbreviate
\be
b_s(z)=b(z)\, e^{2\phi(z)/3}=b(z)\lambda^{2/3}(z),\quad b_{s*}=b_s(z_*).
\ee

\begin{figure}[!tb]
\begin{center}

\includegraphics[width=0.45\textwidth]{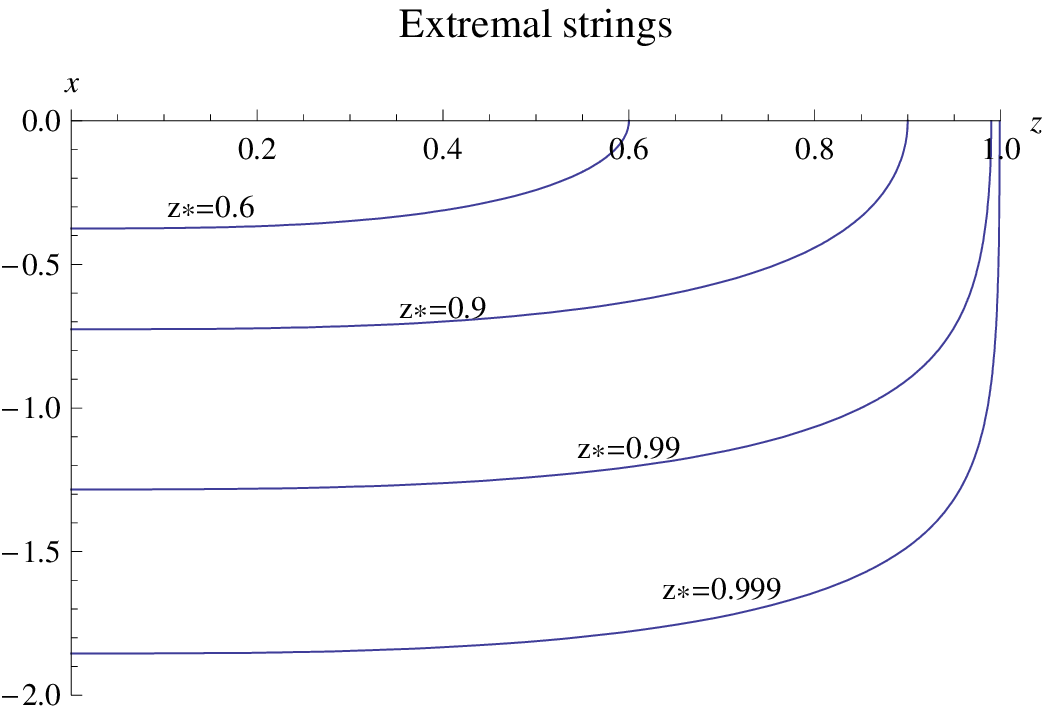}\hfill
\includegraphics[width=0.45\textwidth]{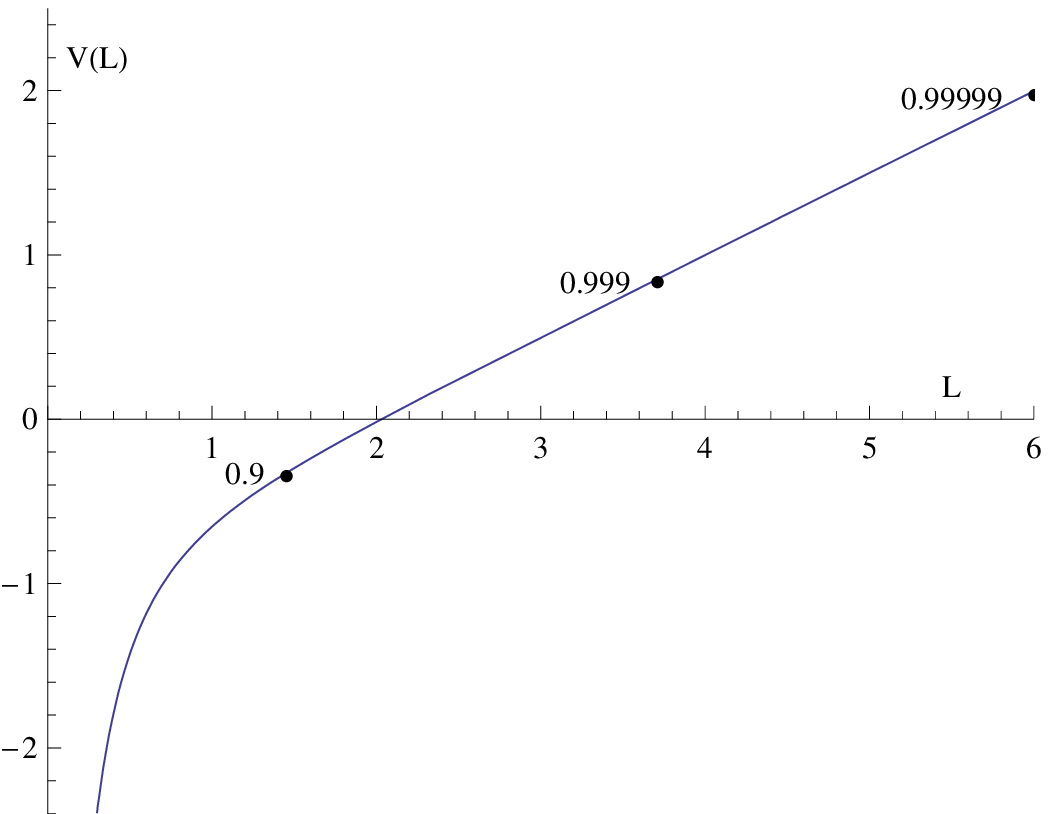}
\end{center}

\caption{\small Left panel: Extremal string configurations in a conformal theory $\phi=0$
computed from \nr{lzstar} for $z_h=1$
and for $z_*=0.6,0.9,0.99,0.999$. Right panel: The potential $V(L)$ as computed from
\nr{lzstar} and \nr{vzstar}. Three values of $z_*$ needed to reach that particular value
of $L$ are marked. }
\label{configs}
\end{figure}

1. Space-space loop:

The relation between $L$ and $z_*$ is given by
\be
L=2\int^{z_*}_0 {dz\over\sqrt{f(z)[b_s^4(z)/b_{s*}^4-1]}}=L(z_*)
\la{lzstar}
\ee
and the potential is
\ba
\hspace{-0.5cm}
V(L)&=&{1\over\pi\alpha'}\int_\epsilon^{z_*} dz
\,b_s^2(z){1\over\sqrt{f(z)[1-b_*^4/b_s^4(z)]}}\la{vzstar}\\
&=&{b_{s*}^2\over 2\pi\alpha'}\,L+{b_{s*}^2\over \pi\alpha'}\int_\epsilon^{z_*}dz
\left[{1\over\sqrt{f(z)}}\sqrt{{b_s^4(z)\over b_{s*}^4}-1}-{b_s^2(z)\over b_{s*}^2}\right]+
{1\over \pi\alpha'}\int_\epsilon^{z_*}dz\,b_s^2(z).
\la{vzstar2}
\ea
In \nr{vzstar2} the first term separates a term proportional to $L$, the second term is finite
when $L\to\infty$ and $z\to0$ (one can put $\epsilon=0$ in it) and the third
term contains the singularity when $z\to0$; its divergence is cancelled when the energy of two
independent (anti)quarks is included \cite{kinar}.
Numerical examples are given in Fig.~\ref{configs} for $z_h=1$.

One observes from Fig.~\ref{configs} that the large-$L$ domain is reached for $z_*\to z_h$.
To see this analytically, expand the integrand in \nr{lzstar}
near $z=z_h$: $f(z)=\dot f(z_h)(z-z_h)+..$, $b_s(z)=b_s(z_*)+\dot b_s(z_*)(z-z_*)+..$ with
$z_*=z_h(1-\epsilon)$. Then
\be
L=\sqrt{{b_s(z_h)\over \dot f(z_h)\dot b_s(z_h)}}\log{4\over 1-z_*/z_h}.
\ee
Thus $L$ diverges when $z_*\to z_h$ provided that $\dot b_s(z_h)<0$
(always $\dot f(z_h)<0$ at $T>0$). Since
$\dot b_s=b_s\fra23\dot\lambda/\lambda\cdot[1+3\lambda/(2\beta)]<0$ the inequality demands
$\beta(\lambda_h)>-\fra32\lambda_h$. This is true as is seen from Figs. \ref{fig_thermalbeta}
and \ref{fig_thermalbetasmallT}, even down to $T=0.96T_c$.
From \nr{vzstar2} with $z_*\to z_h$ the final result is
\be
\sigma_s={1\over 2\pi\alpha'}b^2(z_h)\lambda^{4/3}(z_h).
\label{result1}
\ee
This is the $g_{xx}$ component of the metric \nr{ansatz} in the string frame \nr{stringeinstein}
multiplied by the string tension $1/(2\pi\alpha')$ \cite{kinar}.

2. Time-space loop.

In this case the relations are ($f_*\equiv f(z_*)$)
\be
L=2\int^{z_*}_0 {dz\over\sqrt{f(z)[b_s^4(z)f(z)/(b_{s*}^4f_*)-1]}}=L(z_*)
\la{ltstar}
\ee
and
\ba
\hspace{-0.5cm}
V(L)&=&{1\over\pi\alpha'}\int_\epsilon^{z_*} dz \,b_s^2(z)
{1\over\sqrt{1-b_{s*}^4f_*/[b_s^4(z)f(z)]}}\la{vtstar}\\
&=&{b_{s*}^2\sqrt{f_*}\over 2\pi\alpha'}\,L+{b_{s*}^2\over \pi\alpha'}\int_\epsilon^{z_*}dz
\sqrt{{f_*\over f(z)}}
\sqrt{{b_s^4(z)f(z)\over b_{s*}^4f_*}-1},
\la{vtstar2}
\ea
where the $z\to0$ divergence in $V$ can be regulated as for an $x,y$ loop. If $f\not=1$ $L$ does not
diverge, the equilibrium state is that of two separate quarks \cite{yee,kty}. However,
taking $f=1$ and expanding near $z=z_*$ one finds that $L=(z_*b_{s*}/[-\dot b_s(z_*)])^{1/2}$ so that
$L$ will diverge for that value of $z_*=z_\rmi{min}$ satisfying
\be
\dot b_s(z_\rmi{min})=0.
\ee
From \nr{vtstar2} the string tension then is given by
\be
\sigma={1\over 2\pi\alpha'}b^2(z_\rmi{min})\lambda^{4/3}(z_\rmi{min}).
\la{sigma}
\ee
The functions $b(z),\,\lambda(z)$ in \nr{result1} and \nr{sigma} are different, in the former they
are computed for $f(z)\not=1, \,f(z_h)=0$, in the latter $f(z)=1,\,z_h\to\infty$.

\subsection{Evaluating the result for spatial string tension}
The lattice results for $\sigma_s$ in finite $T$ SU(3) Yang-Mills we want to compare with \cite{boyd}
are measured for $T_c<T<4.5\,T_c$. The QCD computation in Section \ref{sect2} should be a good
estimate of the data even at higher temperatures, the better the larger $T$ is, see Fig.~\ref{fig4}.

It is rather automatic to evaluate the expression \nr{result1} numerically \cite{nitti}, but to
control various effects contributing to it we want to see how far we can get with the approximate
potential \nr{potappro} and the approximate beta function
$\beta(\lambda)=-\beta_0\lambda^q$ in \nr{betaappro}.
The overall normalisation will be fixed from the $T=0$ string tension
$\sigma\approx(440\,{\rm MeV})^2$, measured as $T_c/\sqrt{\sigma}$ \cite{teper1}, using \nr{sigma}. This
has the important property that the arbitrary normalisation of $\lambda$ cancels.

The result \nr{result1} can be analytically approximated as follows:
\ba
{T\over\sqrt{\sigma_s}}&=&\sqrt{2\pi\alpha'}\,{T\over b(z_h)}\,{1\over\lambda^{2/3}(z_h)}\\
&=&\sqrt{2\pi\alpha'}{1\over \pi\CL\left[1-\fra{4}{9(q-1)^2}\log^{-1}\fra{\pi T}{\Lambda}\right]}
\left[(q-1)\beta_0\log\fra{\pi T}{\Lambda}\right]^{2/(3q-3)}\la{2ndres}\\
&=&\sqrt{2\pi\alpha'}{1\over \pi\CL\left[1-\fra{4}{49}\log^{-1}{\pi T\over\Lambda}\right]}
\left[1140\log\fra{\pi T}{\Lambda}\right]^{2/7}\\
&=&\sqrt{2\pi\alpha'}{1\over \pi\CL\left[1-\fra{4}{49}\log^{-1}\fra{0.88T}{T_c}\right]}
\left[1140\log\fra{0.88T}{T_c}\right]^{2/7},
\ea
where we first used the small-$z$ expansions in the Appendix on the leading log level,
writing in the arguments of the logarithms $b(z_h)/b_0=b(z_h)/(\CL\Lambda)=1/(\Lambda z_h)=\pi T/\Lambda$,
then introduced the explicit values $q=10/3,\,\beta_0=489$ used in the numerics \cite{kiri4} and
finally and most subtly, used $0.88\Lambda=\pi T_c$ from the numerics.

To get the normalisation we shall use the lattice data for $T_c/\sqrt{\sigma}=0.597+0.45/N_c^2$
\cite{teper1} and its prediction \nr{sigma} in this model. To have an analytic approximation for
$b_s(z_\rmi{min}),\,\,\dot b_s(z_\rmi{min})=0$ we again use the
beta function $\beta(\lambda)=-\beta_0\lambda^q$,
although the accuracy deteriorates with increasing $\lambda$. Since $\dot b_s=0$ implies
$\beta(\lambda)+\fra32\lambda=0$ one immediately obtains
$\lambda(z_\rmi{min})=(3/(2\beta_0))^{1/(q-1)}$ and then from
Eq.\nr{betalam} $b(z_\rmi{min})=b_0\exp[2/(3(q-1)]$. Thus, with $b_0=\CL\Lambda$,
\be
 b_s(z_\rmi{min})=\CL\Lambda\left({3e\over 2\beta_0}\right)^{2/(3(q-1))}.
 \la{bsminappro}
\ee
Numerical computation ($\CL=1,\,\Lambda=1/174$) with the unapproximated potential \nr{pot}
gives $b_s(z_\rmi{min})=0.00113$; \nr{bsminappro} with parameter values \nr{paramvalues}
gives 0.00146. From \nr{sigma} then
\be
\sqrt{2\pi\alpha'}=\pi\CL {T_c\over\sqrt{\sigma}}\,{\Lambda\over\pi T_c}\,
\left({3e\over 2\beta_0}\right)^{2/(3(q-1))}.
\la{2pia}
\ee
With the lattice result $T_c/\sqrt{\sigma}=0.6$, the numerical result $\pi T_c=0.88\Lambda$,
the value of $\beta_0$
and the correct numerical value of $b_s(z_\rmi{min})$ (reduction of the RHS of \nr{2pia} by
0.00113/0.00146) this can be converted to
\be
{\sqrt{\alpha'}\over \CL}\equiv {\ell_s\over\CL}={1\over6.0}.
\ee
This result depends on the arbitrary normalisation of $\lambda$.

Combining \nr{2ndres} and \nr{2pia} $\beta_0$ cancels and the final result is
\ba
{T\over\sqrt{\sigma_s}}&=&{T_c\over\sqrt{\sigma}}\,{\Lambda\over\pi T_c}\,\,
{1\over 1-\fra{4}{9(q-1)^2}\log^{-1}\fra{\pi T}{\Lambda}}\,\,
\left[\fra32 e(q-1)\log\fra{\pi T}{\Lambda}\right]^{2/(3q-3)}
\\&=&
{T_c\over\sqrt{\sigma}}\,{1.14\over 1-\fra{4}{49}\log^{-1}\fra{0.88 T}{T_c}}\,\,
\left[\fra72 e\log\fra{0.88 T}{T_c}\right]^{2/7}.
\la{finres}
\ea

\vfill
\begin{figure}[!b]
\begin{center}

\includegraphics[width=0.6\textwidth]{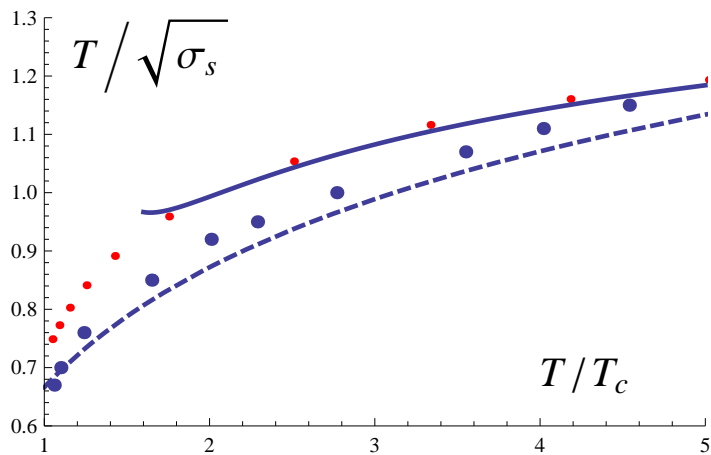}
\end{center}

\caption{\small $T/\sqrt{\sigma_s}$ for the range of $T/T_c$
measured on the lattice \cite{boyd}. The thick points are the lattice data,
the dashed curve is the QCD prediction \nr{tovers}, the smaller points are the result
of numerical integration of the gravity equations of this model and the
continuous curve is the leading log approximation \nr{finres} thereof. The last
can meaningfully be extended down to $T=1.6T_c$.
}
\label{fig3}
\end{figure}


\begin{figure}[!b]
\begin{center}

\includegraphics[width=0.6\textwidth]{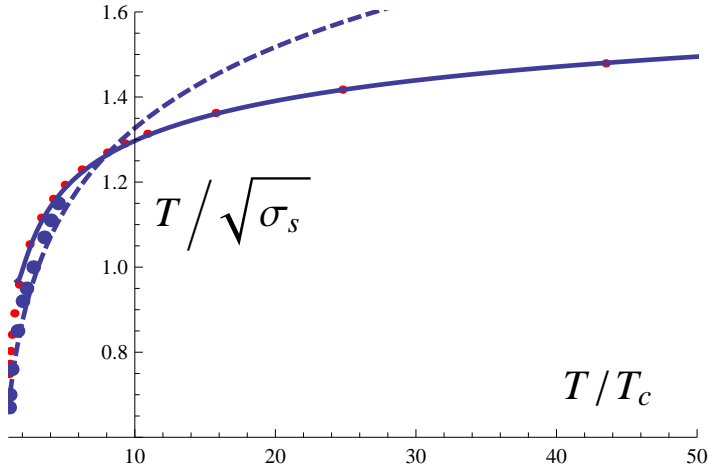}
\end{center}

\caption{\small As the previous figure (which blows up the left part of this figure)
but for values of $T/T_c$ up to 50. Note how the QCD result (dashed line) grows faster with $T$.}
\label{fig4}
\end{figure}

A comparison of the model with lattice data is shown in Figs. \ref{fig3} and \ref{fig4}. We comment
on the result as follows:
\begin{enumerate}
\item The analytic leading log approximation \nr{2ndres} agrees excellently with the
numerical computation using the full potential \nr{pot}
for $T\gsim1.6T_c$ provided that one
supplements it with the numerical result $0.88\Lambda=\pi T_c$.
\item The parameter $\beta_0=\fra{21}8V_1\sqrt{V_3}$ cancelled in \nr{finres}, but if the
full potential \nr{pot} is used, there may be some dependence on it for $T_c<T\lsim 1.6T_c$.
\item The QCD result \nr{tovers} was given as a numerically computed $T$-independent 3d quantity times
a perturbatively computed $T$ dependence. In analogy, here the result is given as the numerically
computed 4d quantity $T_c/\sqrt{\sigma}$ times a $T$ dependence computed from a gauge-gravity duality
model.
\item Within the range where lattice data exists, it is fitted well by both by the QCD result and
the AdS/QCD model discussed here. For larger $T$ the QCD prediction varies more rapidly
with $T$, $\sim\log T/T_c$, than the AdS/QCD model, $\sim \log^{2/7}(T/T_c)$. The exponent here
is $2/(3(q-1))$ with $q=10/3$, as followed from the small-$\lambda$ expansion of the
potential \nr{pot}. In the extreme UV one meets with the QCD beta function with $q=2$,
corresponding to the result behaving $\sim \log^{2/3}(T/T_c)$. If the QCD-like behavior persists,
the present AdS/QCD model will require some modification for $T\gsim 5T_c$.
\end{enumerate}

\section{Conclusions}         
\label{sec:conclusions}
We have in this article considered a particularly simple quantity in finite $T$
quarkless QCD, the spatial string tension, both in QCD and in a model for
gauge/gravity duality, AdS/QCD. The QCD result is perfectly under control:
after one nonperturbative number is determined numerically, the rest follows analytically
via symbolic perturbative computations, which may be technically very demanding.
For the AdS/QCD model
the result needs solutions of Einstein's gravity equations and a particular
contribution of this article has been developing analytically tractable approximations
to the model in \cite{kiri3}.

There are some interesting parallels and differences
in the QCD and AdS/QCD results. Both start
from a $T$ independent nonperturbative quantity,
for QCD the 3 dimensional Yang-Mills string tension,
for AdS/QCD from a 4 dimensional computation of the $T=0$ string tension,
expressed as $T_c/\sqrt{\sigma}$. The $T$ dependence in QCD comes
from perturbatively reintroducing the scales $gT$ and $\pi T$, in AdS/QCD from the
extra dimensional coordinate dependence of the metric and dilaton in the
model used to break conformal invariance.

In the range where lattice data exists, Fig. \ref{fig3}, both calculations
agree with the data within reasonable estimates of errors. At higher $T$, Fig. \ref{fig4},
the AdS/QCD result, with $q=10/3$ used here, increases more slowly than the QCD result
and it should be possible to distinguish between the two.

There are puzzling features in both approaches. The QCD approach adds perturbative
corrections to a nonperturbative number and it is unexpected that it works so well
down to $T_c$. The AdS/QCD model fits very well and elegantly bulk thermodynamics \cite{kiri4}
using the potential \nr{pot}, 
but the power of $\log T$ is quite different from that in 
the QCD running coupling \nr{g2op}.
This is not surprising near $T_c$ but one would expect that at least at some high $T$
the matter would probe distances small enough to see QCD asymptotic freedom.

The spatial string tension has also been measured for $N_f=2+1$ QCD \cite{cheng}.
An interesting further project would be to study this in AdS/QCD by developing
the AdS/QCD model used here to also include fundamental quarks.

\vspace{1cm}
{\it Acknowledgements}.
We thank E. Kiritsis, Mikko Laine, York Schr\"oder and Patta
Yogendran for discussions and
correspondence. We are particularly indebted to Francesco Nitti for the Mathematica code
used to solve the Einstein equations of this model and for advice concerning its use.

\appendix

\section{Appendix}
We summarize here the solution of Eqs.\nr{eq1}-\nr{crucial} for the beta function
\nr{betaappro}, $\beta(\lambda)=-\beta_0\lambda^q$. First, from Eq.\nr{crucial},
\be
Q\equiv\log{b\over b_0}={1\over(q-1)\beta_0\lambda^{q-1}},
\la{betalam}
\ee
where $b_0$ is a constant, the analogue of $\Lambda_\rmi{QCD}$, the scale at which
$\lambda$ diverges. From \nr{W}
\be
W={1\over\CL}\exp{4\over9(q-1)^2\log(b/b_0)}.
\ee
The second equation can then be written in the form
\be
{d\lambda\over dz}={db\over dz}{\beta\over b}=-\beta(\lambda)b(\lambda)W(\lambda),
\la{zlab}
\ee
from which by integration
\be
z={\CL\over b_0}\sum_0^\infty {1\over n!}\left(-{4\over9(q-1)^2}\right)^n \Gamma(1-n,Q),
\ee
where
\be
\Gamma[1-n,Q]=\int_Q^\infty dt\,e^{-t}t^{-n}=e^{-Q}{1\over Q^n}
\left[1-{n\over Q}+{n(n+1)\over Q^2}-..\right].
\label{gammaexp}
\ee
Correct dimensions are here given by
\be
\Lambda={b_0\over\CL}.
\ee
Using \nr{zlab} one can replace $z$ as a fifth coordinate by $b$, $\lambda$ or $Q$.
For example,
\ba
\int_0^z{d\bar z\over b^3(\bar z)}&=&{\CL\over b_0^4}\int_Q^\infty dy\exp
\left[-4y-{4\over9(q-1)^2y}\right]\nonumber \\ &=&
{\CL\over4b_0^4}
\sum_0^\infty {1\over n!}\left(-{16\over9(q-1)^2}\right)^n \Gamma(1-n,4Q).
\ea
If the potential satisfies the constraint \nr{V}, there is a horizon with the temperature
\ba
{1\over\pi T}&=&{\CL\over b_0}e^{3Q}\int_{4Q}^\infty dy \exp[-y-16/(9(q-1)^2y)]\nn
&=&
{\CL\over b_0^4}
\sum_0^\infty {1\over n!}\left(-{16\over9(q-1)^2}\right)^n \Gamma(1-n,4Q(z_h))\,\,b^3(z_h).
\la{teequu}
\ea
Using \nr{gammaexp} one can derive various small-$z$
($z\ll1/\Lambda$, small $\lambda$, large $Q$) approximations.
One has, for example,
\ba
b(z)&=&{\CL\over z}\left[1+{4\over9(q-1)^2\log(\Lambda z)}+
{4(2+9(q-1)^2)\over 81(q-1)^4\log^2(\Lambda z)}+...\right],\\
{1\over \lambda^{q-1}}&=&(q-1)\beta_0\log{1\over\Lambda z}
\left[1-{4\over 9(q-1)^2\log^2(\Lambda z)}+{4(2+7(q-1)^2)\over 81(q-1)^4\log^3(\Lambda z)}+...\right],\\
{1\over\pi T}&=&z_h\left[1+{1\over 3(q-1)^2\log^2(\Lambda z_h)}+..\right]\\
&=&{\CL\over b(z_h)}\left[1-{4\over9(q-1)^2\log[b(z_h)/b_0]}+...\right].\la{teebee}
\ea
Finally, as in \cite{kiri3,kiri4} one can obtain the entire bulk thermodynamics by integrating
$s(T)=p'(T)$, $s(T)=b^3(z_h)/(4G_5)$. 
%
%
%
%
%
%

\end{document}